# Cerium oxide catalyzed disproportionation of hydrogen peroxide: a closer look at the reaction intermediate

**Giusy Finocchiaro[1,2]\*, Xiaohui Ju[3]\*, Braham Mezghrani[1] and Jean-François Berret[1]\***

*[1]Université Paris Cité, CNRS, Matière et systèmes complexes, 75013 Paris, France*

*[2]Institute of Photonics and Electronics of the Czech Academy of Sciences, Chaberská1014/57, 182 51 Prague, Czech Republic*

*[3]Center for Nanorobotics and Machine Intelligence, Department of Chemistry and Biochemistry, Mendel University in Brno, Zemedelska 1, 613 00 Brno, Czech Republic*

**Abstract**: Cerium oxide nanoparticles (CNPs) have recently gained increasing interest as redox enzyme-mimetics to scavenge the intracellular excess of reactive oxygen species, including hydrogen peroxide ($H_2O_2$). Despite the extensive exploration of CNP scavenging activity, there remains a notable knowledge gap regarding the fundamental mechanism underlying the CNP catalyzed disproportionation of $H_2O_2$. In this Letter, we present evidence demonstrating that $H_2O_2$ adsorption at CNP surface triggers the formation of stable intermediates known as cerium-peroxo complexes ($Ce-O_2^{2-}$). The cerium-peroxo complexes can be resolved by Raman scattering and UV-Visible spectroscopy. We further demonstrate that the catalytic reactivity of CNPs in the $H_2O_2$ disproportionation reaction increases with the Ce(III) fraction. The developed approach using UV-Visible spectroscopy for the characterization of $Ce-O_2^{2-}$ complexes can potentially serve as a foundation for determining the catalytic reactivity of CNPs in the disproportionation of $H_2O_2$.

Cerium(IV) oxide, also known as cerium oxide ($CeO_2$) exhibits redox reactivity due to the facile valence switching between Ce(IV) and Ce(III) oxidation states.[1,2] This dynamic interplay between oxidation states is closely linked to the presence of oxygen vacancies within the lattice structure. In the past few years, there has been a growing interest utilizing cerium oxide nanoparticles (CNPs) as redox enzyme-mimetics in the field of nanomedicine.[3-8] These CNPs have shown remarkable enzyme-mimetic properties at physiological conditions,[9] making them potential cytoprotective agents against reactive oxygen species (ROS).[4,7,8,10-13] In particular, CNPs have demonstrated the ability to scavenge superoxide ($O_2^-$) and hydrogen peroxide ($H_2O_2$), which are physiologically degraded by the redox enzymes superoxide dismutase (SOD) and catalase (CAT), respectively.[3,7,14] However, the debate about the relationship between the catalytic disproportionation of $H_2O_2$ into $H_2O$ and $\frac{1}{2}O_2$ and CNP scavenging activity, particularly in relation to their physico-chemical properties remains unresolved. This lack of consensus poses challenges for the definitive conclusion and application of CNPs *in vitro* and *in vivo*.[7,14-18] It has been generally accepted that Ce(III) is associated with the enhanced activity of SOD-mimetics, while Ce(IV) is responsible for the CAT-mediated disproportionation of $H_2O_2$.[3,7,19] This understanding has long served as a fundamental framework for investigating CNP catalytic mechanisms in various applications.

However, it is important to note that there are differing viewpoints on this matter. Through a systematic investigation of the size dependent SOD- and CAT-mimetic activity of CNPs, Baldim *et al.*[14] observed a correlation between CNP decreasing sizes and the increase of both SOD- and CAT-mimetic activities. Since smaller CNPs exhibit a higher Ce(III) fraction[7,14,20] these results suggest a close relationship between the Ce(III)-content and CNP catalytic activity.[1,7,14,16,18,20] Alongside the previous Ce(III)/Ce(IV) models, Cafun *et al.*[21] proposed an alternative perspective, suggesting that the actives sites responsible for catalyzing $H_2O_2$ disproportionation are not solely interfacial cerium atoms, but rather the electrons that are delocalized within the nanocrystal. Yet, the findings from Cafun *et al.*[21] do not exclude the importance of Ce(III) fraction in enhancing the catalytic activity of CNPs,[14,16,18,20,21] as they also reported higher reactivity of smaller size CNPs catalyzing $H_2O_2$ disproportionation. Furthermore, contrary to the prevailing idea that Ce(III) solely exists at CNP-solvent interface, Hao *et al.* have recently demonstrated the presence of Ce(III)-cations within the core of sub-10 nm nanocubes.[1] This distribution is attributed to the nano-size effect induced by the overall lattice expansion promoting oxygen vacancy formation.

In this letter, we take a closer look at the reaction mechanism of $H_2O_2$ disproportionation catalyzed by CNPs, and provide concrete evidence supporting the hypothesis that the reaction leads to the formation of cerium-peroxo complexes (Ce-$O_2^{2-}$) at the CNP surface as reaction intermediates. While the presence of Ce-$O_2^{2-}$ complexes was proposed by various authors, [17,22-27] this letter represents the first instance of identifying the signatures of cerium-peroxo complexes in CNP-$H_2O_2$ dispersions as a function of the Ce(III) fraction. In our study, we employ complementary spectroscopic techniques, Raman scattering[22,23,25,26,28] and UV-Visible spectroscopy[13,14,16,17,21,26] to resolve the structure and characteristics of the Ce-$O_2^{2-}$ complexes.

By considering two critical parameters, the $H_2O_2$ concentration and the CNP size, the latter being linked to the Ce(III) fraction, we have successfully showcased the possibility of Ce-$O_2^{2-}$ complexes quantification. Therefore, in addition to identifying the CNP-$H_2O_2$ reaction intermediates, the approach we have developed can potentially serve as a significant foundation for characterizing the CNP catalytic reactivity in $H_2O_2$ disproportionation by spectroscopy.

Extensive evidence has demonstrated that the color of a CNP dispersion undergoes instantaneous changes upon $H_2O_2$ addition.[13,15,16,29] This phenomenon has been attributed to the reversible oxidation of Ce(III) to Ce(IV) ions, which serves as a mediator in the $H_2O_2$ disproportionation reaction.[15,16,29] Fig. 1a displays the color changes observed in CNPs with 7.8 nm diameter (referred to as CNP8)[30,31] from yellow to garnet in response to $H_2O_2$ addition at concentrations ranging from 0 to 2000 mM.[14] Further incubation of the samples over a period of one month revealed that the dispersions returned to their original color (**S1**). Following the $H_2O_2$ addition to an 8 g $L^{-1}$ CNP8 dispersion, the samples were freeze-dried to obtain powders, which were then subjected to analysis by Raman scattering. The characteristic Raman bands of pristine

CNP8, identified as bands **(1)** to **(4)** in Fig. 1b, have been widely documented in previous studies.[28,32] The 460 cm[-1] band labelled **(1)** in the figure corresponds to the $F_{2g}$ mode of $CeO_2$ fluorite phase[22,25,28,32] whereas that at 590 cm[-1] **(2)** arises from defects in $CeO_2$ crystal structure, such as oxygen vacancies.[22,25,26,28] The two bands at 730 and 1050 cm[-1] **(3)** stem from nitrate counterions arising from nitric acid added during synthesis.[33,34] When exposed to hydrogen peroxide, a distinctive band **(4)** appears at 840 cm[-1], which could potentially be associated with either the dioxygen stretching vibration of the adsorbed peroxide species $O_2^{2-}$ at the CNP8 surface or residual $H_2O_2$ peroxide.[22,23,35]

To further clarify the origin of the newly observed Raman band **(4)**, a comparative analysis was performed between a 9.8 M $H_2O_2$ solution and a 110 g L[-1] CNP8-$H_2O_2$ dispersion, with $[H_2O_2]$ = 5 M. The findings are depicted in Fig. 1d. While the $H_2O_2$ solution exhibits a sharp band (labelled **(5)** in the figure) at 875 cm[-1],[36] the CNP8-$H_2O_2$ dispersion shows a broader band at 840 cm[-1] with a shoulder extending to higher wavenumbers, that partially overlaps with the 875 cm[-1] dioxygen stretching vibration from $H_2O_2$. As the overlap due to residual $H_2O_2$ in the liquid phase may lead to ambiguous interpretation, we opted to collect Raman data from samples in their powder state.

In the field of heterogeneous catalysis, the activation of dioxygen ($O_2$) at metal oxide interfaces has been extensively investigated and is commonly described by the Mars-van Krevelen mechanism.[24] According to this model, the initial uptake of $O_2$ molecules leads to the formation of metal-peroxo complexes, followed by the catalytic reaction and the subsequent release of the products.[24,25] Similarly here, Ce-$O_2^{2-}$ complexes are formed at the CNP surface upon $H_2O_2$ uptake, as intermediates in the $H_2O_2$ catalytic disproportionation. Previous studies have suggested that intermediate peroxide species can adopt two different bonding configurations at the CNP surface (Fig. 1c).[22-24,37] In one configuration, oxygen atoms bind to two adjacent cerium atoms forming the non-planar bridging μ-peroxo-complex, and corresponding to a Raman shift between 875 cm[-1] and 890 cm[-1]. In the other configuration, both oxygen atoms coordinate to the same cerium atom, forming the side-on η[2]-peroxo-complex[22,26,37] whose Raman band is centered around 831 cm[-1].[22,25,32,35] Regarding the newly observed Raman band (4) in Fig. 1b, its position at 840 cm[-1] indicates that the cerium-peroxo complexes formed at the CNP8 surface can exist in both configurations, with a higher prevalence for the η[2]-peroxo-complexes. Note that the Raman band from the Ce-$O_2^{2-}$ complexes was also observed in the CNP5 dispersions at the same wavelength of 840 cm[-1] (S2). Fig. 1e displays the Raman spectra of CNP8 in powder form obtained one hour after addition of $H_2O_2$ ranging from 2 to 2000 mM, and corresponding to the samples in Fig. 1a. The spectrum of pristine CNP8 is also included for comparison. As the $H_2O_2$ concentration increases, the relative intensity of the Raman band associated with Ce-$O_2^{2-}$ complexes also increases. This observation further suggests a correlation between the hydrogen peroxide concentration and Ce-$O_2^{2-}$ complex formation. The same CNP8-$H_2O_2$ dispersions were prepared again and left to incubate for one month at room temperature. Afterward, they underwent freeze-drying. The comparison of Figs. 1e- and 1f reveals that the Raman band corresponding to the Ce-$O_2^{2-}$ complexes at 840 cm[-1] is no longer observed in the CNP8-$H_2O_2$ spectra after one month of incubation. This observation is consistent with previous findings regarding the color changes observed in CNP8-$H_2O_2$ dispersions, which then revert to their original color (Fig. 1a). It is assumed that in either case the disproportionation of $H_2O_2$ catalyzed by the CNPs is complete.

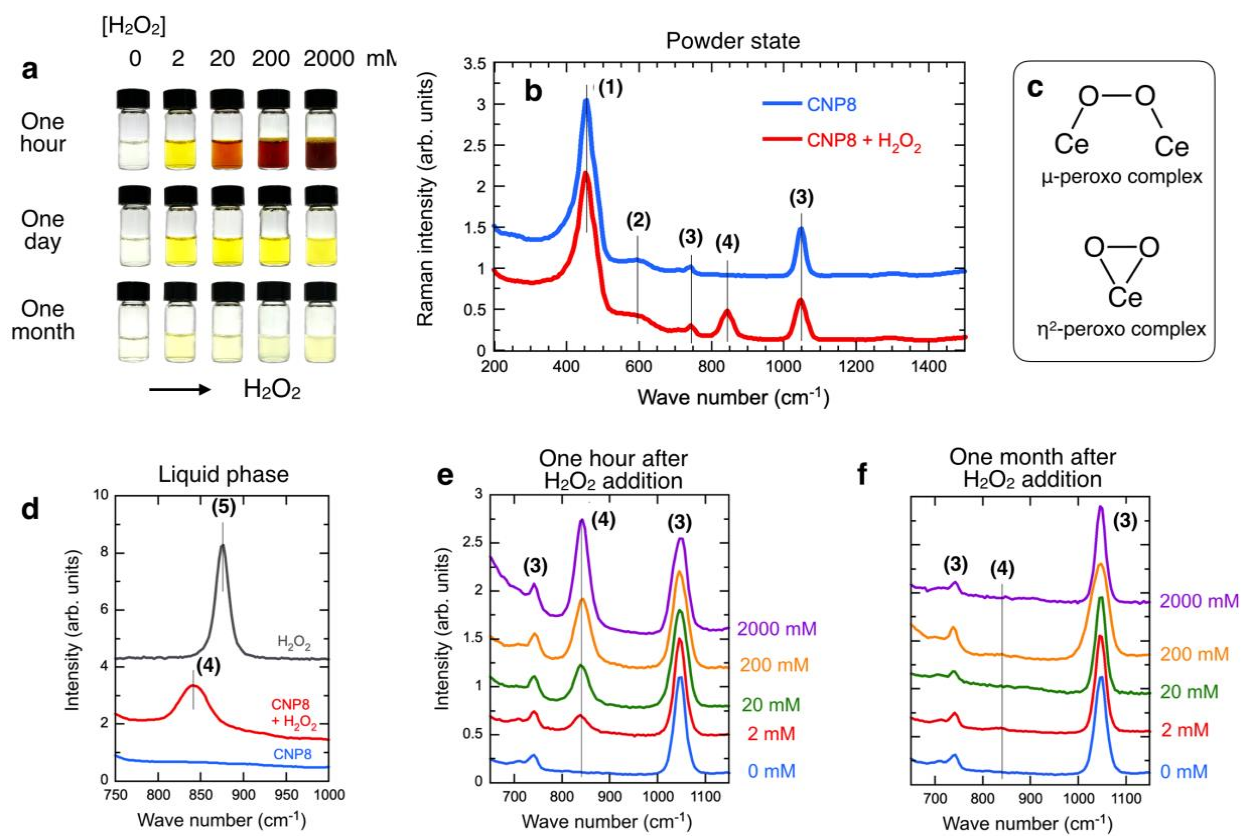

**Figure 1**: **a)** Images of vials containing 8 g L$^{-1}$ CNP8 dispersions with increasing concentrations of hydrogen peroxide (0, 2, 20, 200 and 2000 mM). The pictures were taken one hour, one day and one month after preparation. **b)** Raman spectra of CNP8 and CNP8 after H$_2$O$_2$ addition, recorded between 200 cm$^{-1}$ and 1500 cm$^{-1}$. The samples were in powder form and prepared according to the protocol described in **S3**. **c)** μ-peroxo- and η$^2$-peroxo-complexes formed upon H$_2$O$_2$ adsorption to the CNP surface.[22,26,37] **d)** Raman spectra displaying the stretching vibration of dioxygen between 750 cm$^{-1}$ and 1000 cm$^{-1}$ in the liquid phase for 110 g L$^{-1}$ CNP8 dispersions with and without H$_2$O$_2$. For the H$_2$O$_2$ solution and the CNP8-H$_2$O$_2$ dispersion, the hydrogen peroxide concentration was 9.8 M and 5 M, respectively. **e)** Raman spectra of CNP8-H$_2$O$_2$ samples in powder form. The dispersions were formulated as in **Fig. 1a**, and then freeze-dried shortly after H$_2$O$_2$ addition. **f)** As in **Fig. 1e**, for dispersions that were freeze-dried one month after H$_2$O$_2$ addition. In the two last figures, the Raman spectra were normalized to the nitrate band (3) at 1050 cm$^{-1}$.

It has been hypothesized that the color changes observed in Fig. 1a can be described in terms of a red-shift in UV-Visible absorbance spectra.[15,16,38] Lee *et al.* have proposed a spectrophotometric method for quantifying the CNP antioxidant capacity, which involves monitoring the red-shift of the wavelength at a fixed absorbance.[16] Based on this criterion, the stronger the red-shift, the higher the antioxidant properties. This approach has subsequently been adopted by numerous authors.[13,14,19,39,40] Damatov and Mayer proposed an alternative treatment for the UV-Visible spectra[17] and evaluated the difference in absorbance between samples with and without hydrogen peroxide, revealing the emergence of an absorbance doublet at 285 and 380 nm. The observations concerning the latter 380 nm band have also been corroborated in our previous work.[41,42]

In the second part of this letter, we have used the knowledge gained from Raman scattering to interpret the UV-Visible spectra of CNP8-$H_2O_2$ dispersions, enabling a better understanding of the role of Ce-$O_2^{2-}$ complexes in the UV-Visible absorbance. In this study, we conducted UV-Visible spectroscopic investigations taking into account the size effect of CNPs interacting with $H_2O_2$. It is now well established that CNP size plays a key role in Ce(III) fraction, noted here as $f_{Ce(III)}$, and that fraction increases as the size decreases.[1] Based on our previous work,[14] we have carried out a comparative analysis of CNPs with varying diameters. In addition to the CNP8 from Fig. 1, we have examined CNPs with diameters of 4.5 nm, 23 nm, and 28 nm. These particles will be referred to as CNP5, CNP23, and CNP28, respectively. CNP5 are spherical nanoparticles of diameters of 4.5 nm, while CNP8 consists of agglomerates formed by 2.6 nm nanocrystallites[30,31,42] as evidenced by transmission electron microscopy (Figs. 2a-2b). The larger particles, CNP23 and CNP28, exhibit a broad size distribution and possess a polyhedral shape (Figs. 2c-2d). The X-ray photoelectron spectroscopy (XPS) (S4) determined the Ce(III) fractions for CNP5, CNP8, CNP23 and CNP28, yielding $f_{Ce(III)}$ values of 40%, 20%, 9% and 9%, respectively.

Figs. 2e-h depicts the UV-Visible absorption spectra of CNP5, CNP8, CNP23 and CNP28 dispersions with increased amounts of $H_2O_2$. Without hydrogen peroxide, the spectra all show the characteristic peak of $CeO_2$ nanoparticles at the same wavelength, here 288 nm.[43,44] The spectra in Figs. 2e-h were derived from the raw data by subtracting the solvent contribution and normalizing it with the absorbance value of the $CeO_2$ peak maximum (S3). Treated CNP and CNP-$H_2O_2$ spectra are designated $\tilde{A}_{CNP}(\lambda)$ and $\tilde{A}_{CNP+H_2O_2}(\lambda)$, the tilde making reference to the normalization.[42] At lower wavelengths, it is observed that spectra overlap well, except for [$H_2O_2$] = 200 mM and 2000 mM. The absence of the $CeO_2$ characteristic 288 nm peak is attributed to increased $H_2O_2$ absorbance in the UV region, causing a spectral bias, particularly evident after solvent subtraction (Figs. 2e-h, green and yellow curves). Figs. 2e-h demonstrate a noticeable increase in absorbance within the range of 300-600 nm with increasing $H_2O_2$ concentration.[13,41] For CNP5 and CNP8, the data show the absorbance excess even at concentrations below 1 mM, while for CNP23 and CNP28 this excess remains low over the entire $H_2O_2$ range. Figs. 2i-l illustrates the absorbance excess $\tilde{A}_{exc}(\lambda) = \tilde{A}_{CNP+H_2O_2}(\lambda) - \tilde{A}_{CNP}(\lambda)$, revealing the presence of an additional absorbance peak centered around 365 nm.[17] This absorbance excess is found to increase with increasing $H_2O_2$ concentrations. It is slightly blue-shifted, its maximum decreasing from 375 nm to 365 nm with increasing [$H_2O_2$] (S5). Figs. S6 reveal that the 365 nm absorbance peaks of CNP5 and CNP8 coincide and exhibit similar shapes when normalized. The superposition of the 365 nm absorbance peaks, regardless of particle size or hydrogen peroxide concentration suggests that they can be attributed to the same chemical species, indicating a common underlying phenomenon. In contrast, the peaks observed in CNP23 and CNP28 at [$H_2O_2$] = 200 and 2000 mM are located in a lower $\lambda$-range, around 340 nm (S7) and exhibit different shapes. For these reasons, the data will not be excluded from the analysis. Monitoring the CNP8-$H_2O_2$ absorbance peaks at 365 nm over a one-month period reveals an interesting observation: the initial absorbance excess induced by $H_2O_2$ addition gradually disappears over this period (S8). These findings indicate a time-dependent evolution phenomenon that correlates well with the disappearance of the 840 cm$^{-1}$ Raman band assigned to cerium-peroxo complexes (Fig. 1f).

Additionally, they are also consistent with the observed color changes in the CNP8-H₂O₂ dispersions throughout the one-month period (Fig. 1a). Indeed, the above findings strongly indicate a correlation between the 365 nm absorbance peak, the newly identified Raman band, and the observed CNP8-H₂O₂ color changes. These observations collectively provide compelling evidence that this phenomenon can be attributed to the formation and evolution of cerium-peroxo complexes as the intermediates of H₂O₂ catalytic disproportionation mediated by CNPs. Moreover, the absorbance peaks are indicative of a ligand-to-metal charge transfer transition, wherein the peroxide ligand transfers charge to the cerium atom, giving rise to the observed patterns.[17,45-47] Finally, we noticed also a correlation between $f_{Ce(III)}$ and the intensity of 365 nm UV-Visible band, indicating that the presence of oxygen vacancies associated with the Ce(III) fraction play a pivotal role in the catalytic interaction between H₂O₂ and CNPs.

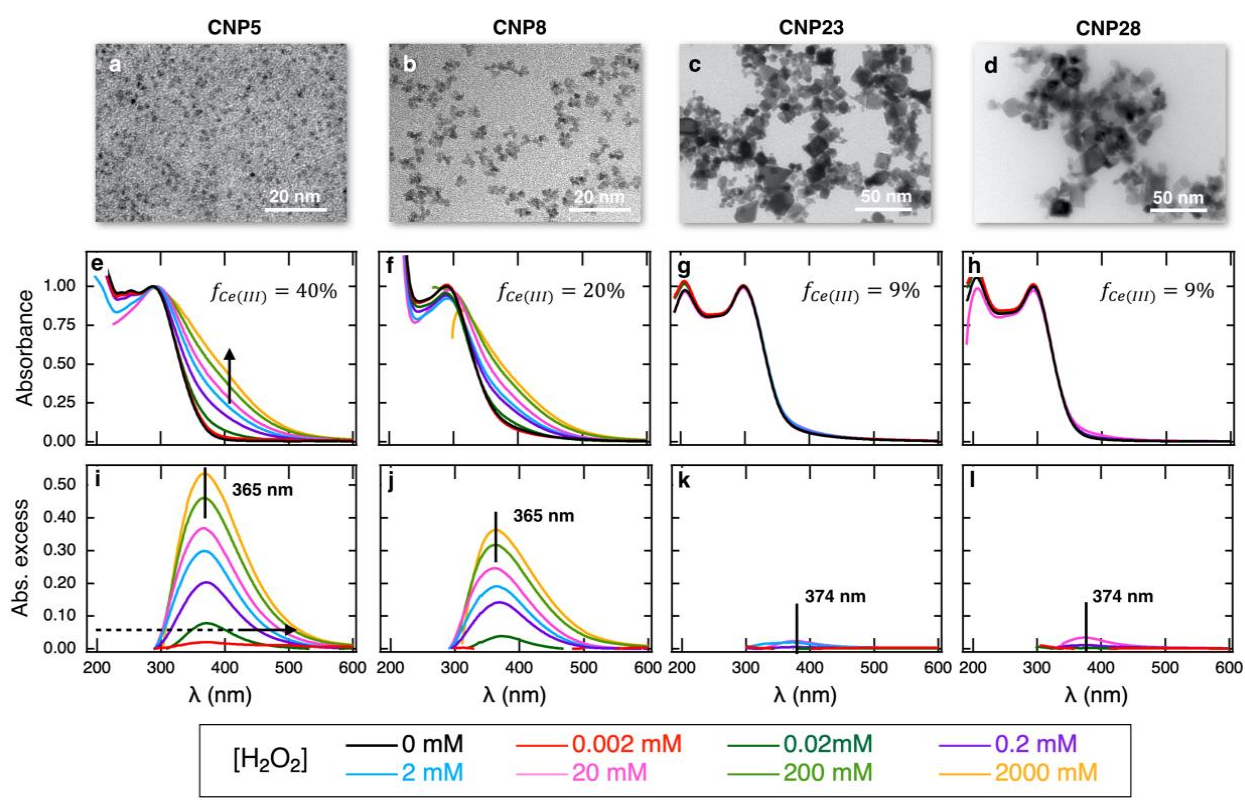

**Figure 2**: **a-d)** Representative transmission electron microscopy images of cerium oxide nanoparticles of sizes 4.5 nm (CNP5), 7.8 nm (CNP8), 23 nm (CNP23) and 28 nm (CNP28), respectively. **e-h)** UV-Visible absorbance spectra of CNP5, CNP8, CNP23 and CNP28, respectively, with H₂O₂ concentration increasing from 0.002 to 2000 mM, the CeO₂ concentration being approximately 0.03 g L⁻¹. The vertical arrow in **e)** highlights the absorbance increase observed at 365 nm. **i-l)** Absorbance excess $\tilde{A}_{exc}(\lambda)$ for CNP5, CNP8, CNP23 and CNP28, respectively. In **i)**, the horizontal arrow at absorbance 0.05 indicates the apparent red-shift observed with increasing hydrogen peroxide.

In light of the widespread utilization of the red-shift method for CNP characterization,[16] we aim to contribute additional insights by addressing the discrepancies between reported interpretations and our own observations. From the data in Figs. 2i-l, it can be seen that with increasing [$H_2O_2$] the absorbance excess $\tilde{A}_{exc}(\lambda)$ increases and gradually extends into the visible region, here between 400 and 500 nm. This effect is illustrated for CNP5 in Fig. 2i by a horizontal arrow. However, it is important to note that the interpretation in terms of red-shift offers a limited picture of CNP/$H_2O_2$ interaction and relies on the assumption that the red-shift can be observed at an arbitrarily fixed absorbance level. In contrast, the approach proposed in this Letter does not rely on such assumptions and has the potential to accurately quantify the cerium-peroxo complexes at CNP surfaces by spectroscopy without additional reagent. Compared to Damatov and Mayer's data, [17] which showed a double peak in the $\tilde{A}_{exc}(\lambda)$ spectrum, one at 285 nm associated with the oxidation of Ce(III) to Ce(IV), and the second at 380 nm originating from peroxo-complexes, our results only feature a single peak, which is furthermore shifted by 15 nm from theirs. This discrepancy has tentatively been ascribed to the different experimental conditions used in Ref.[17], such as the smaller size CNPs (2.4 nm), the coating of CNPs with oleate capping ligands and the polar organic solvent in which the dispersions were prepared. Note finally that the position of the absorbance maximum at 365 nm corresponds well to those observed under similar conditions in our previous work.[41,42]

To assess the size-dependent effect on CNP reactivity towards $H_2O_2$ catalytic disproportion, we plotted the absorbance at 365 nm as a function of $H_2O_2$ concentration, noted hereafter $\tilde{A}_{exc}([H_2O_2])$. The data obtained for CNP5 and CNP8 are shown in Fig. 3a, while those for CNP23 and CNP28 are shown in Fig. 3b. For the latter, and following our discussion in the previous section, the results at [$H_2O_2$] = 200 mM and 2000 mM were excluded. The $\tilde{A}_{exc}([H_2O_2])$ data were fitted with the Hill's equation, a model commonly used to analyze the enzyme-mimetic reactions:[42,48]

$$\tilde{A}_{exc}([H_2O_2]) = \tilde{A}_{max}\frac{[H_2O_2]^\alpha}{c_0^\alpha + [H_2O_2]^\alpha} \qquad (1)$$

where $\tilde{A}_{max}$ is the absorbance excess at saturation, $c_0$ is the concentration of $H_2O_2$ at which $\tilde{A}_{exc}([H_2O_2]) = \tilde{A}_{max}/2$ and $\alpha$ is the Hill coefficient.[42] For $\alpha > 1$, the Hill's model predicts a sigmoidal shape, showing an initial steep rise before reaching a saturation level. Conversely for $\alpha \leq 1$, the curve displays an initial linear ($\alpha = 1$) or sub-linear ($\alpha < 1$) increase, before eventually saturating at higher $H_2O_2$ concentrations,[48,49] as observed in Figs. 3a-3b. From a mechanistic perspective, $\alpha$ represents the extent of cooperativity in the binding process, allowing for an estimation of the number of ligands binding to the same receptor in the case $\alpha > 1$.[48,49] Fig. 3c displays the adjustable parameters $\alpha$, $\tilde{A}_{max}$ and $c_0$ retrieved from the Hill's model. Among the four investigated CNPs, we observe a distinctive adsorption behavior characterized by comparable $\alpha$ parameters, which falls in the range of 0.33-0.40. In the Hill's model, $\tilde{A}_{max}$ relates to the availability of adsorption sites at CNP surfaces for the binding with $H_2O_2$. The parameter $\tilde{A}_{max}$ is found to increase with the Ce(III) fraction, as illustrated in Fig. 3d. This figure suggests that UV-Visible spectroscopy for characterizing cerium-peroxo complexes may serve as a basis for studying the CNP catalytic activity in $H_2O_2$ disproportionation reaction. Pertaining to the parameter $c_0$ which is linked to the dissociation constant of the reaction, we observe that it increases continuously with the CNP size, from 2.6 mM to 15.4 mM, indicating that the $H_2O_2$ affinity for the CNP surface is greater with smaller sizes and higher Ce(III) fractions. Additionally, we estimated the number of moles of $H_2O_2$ needed to saturate one mole of CNP5, CNP8, CNP23 and CNP28, which turned out to be approximately 30, 40, 140 and 150 (S3). These figures exhibit a size-dependent effect, indicating that larger CNPs generally require a higher amount of $H_2O_2$ for saturation. This outcome is again in line with our previous observations[14,42] confirming a consistent trend regarding the reactivity of CNPs to $H_2O_2$ disproportionation and Ce(III) fraction $f_{Ce(III)}$.[14,16,18] Finally, in Fig. 3e, we observe a direct relationship between Raman scattering and UV-Visible spectrometry data for CNP8. There, the 840 cm⁻¹ Raman band intensity from Fig. 1e is displayed as function of the observed UV-Visible absorbance $\tilde{A}_{exc}(\lambda = 365 \text{ nm})$ found for CNP8-$H_2O_2$ dispersions. The comparison reveals a linear dependence between the two intensities, reinforcing the hypothesis that these two results are interconnected and associated with the same phenomenon, that is formation of Ce-$O_2^{2-}$ complexes as intermediates of CNP catalyzed $H_2O_2$ disproportionation.

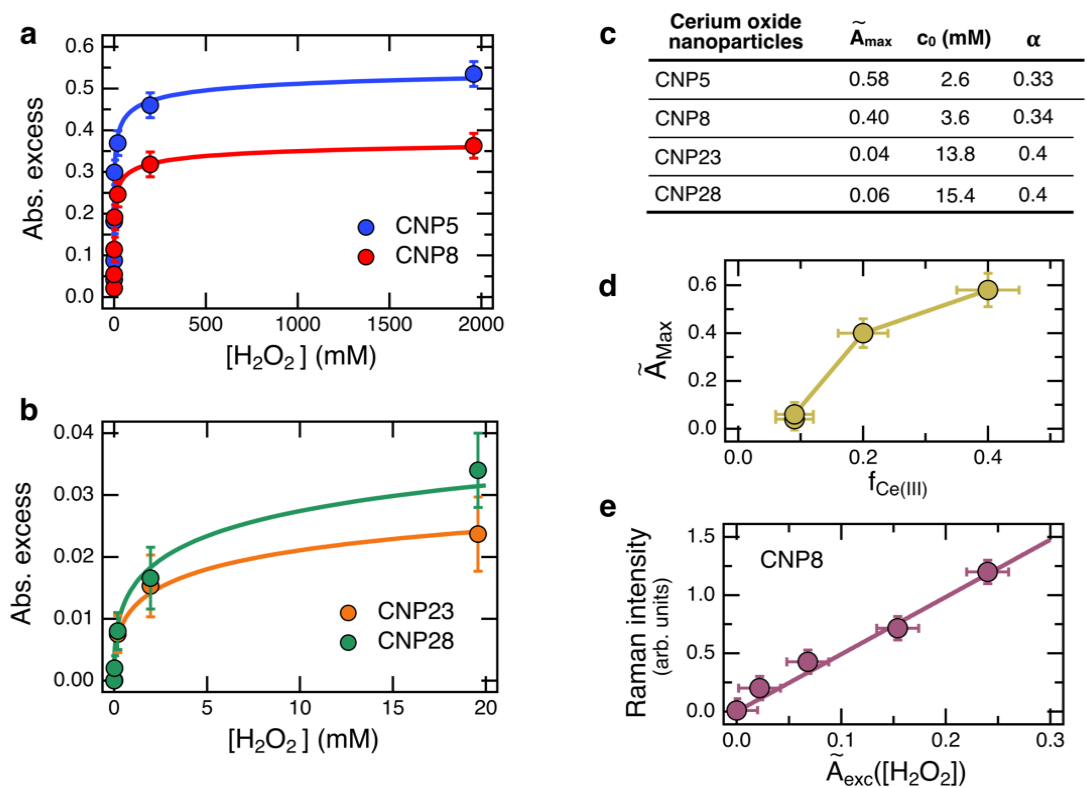

**Figure 3**: **a)** Absorbance excess $\tilde{A}_{exc}(\lambda) = \tilde{A}_{CNP+H_2O_2} - \tilde{A}_{CNP}$ at the wavelength $\lambda = 365$ nm after addition of increasing amounts of $H_2O_2$ and one hour equilibration time for CNP5 and CNP8. The continuous line was obtained from the Hill equation (Eq. 1). **b)** As in a) for CNP23 and CNP28. **c)** Hill parameters used to fit the absorbance excess as a function of the $H_2O_2$ concentration. **d)** Variation of parameter $\tilde{A}_{max}$ as a function of the Ce(III) fraction for CNP5, CNP8, CNP23 and CNP28. **e)** Intensity of the Raman band at 840 cm⁻¹ as function of the absorbance maximum $\tilde{A}_{exc}$ for CNP8-$H_2O_2$ dispersions. The observed linear variation suggests that the band at 840 cm⁻¹ and the 365 nm UV-Visible absorbance peak have the same origin, namely the formation of the peroxo-complexes at the CNP surface.

In this Letter, we present a detailed examination of the mechanism underlying the disproportionation of hydrogen peroxide catalyzed by cerium oxide nanoparticles of different sizes between 4.5 and 28 nm. Our investigation reveals the presence of stable cerium-peroxo complexes, formed at the CNP surface, which serve as reaction intermediates. Using Raman scattering and UV-Visible spectroscopy, we identify the surface complexes as intermediates in the reaction by monitoring their spectroscopic signatures upon $H_2O_2$ addition, and their disappearance upon completion of the disproportionation reaction. Our study demonstrates that the presence of the cerium-peroxo complexes is directly related to a UV-Visible absorbance band centered at 365 nm, in contrast with previous reports suggesting a correlation with the red-shift of the absorbance spectra.[16] Furthermore, we investigate the effect of the Ce(III) fraction on the CNP reactivity in catalyzing the disproportionation of $H_2O_2$. Hydrogen peroxide addition to CNP dispersions induces noticeable changes in the UV-visible and Raman spectra, which become more pronounced with increasing Ce(III) fraction. These findings open up the possibility of quantifying the cerium-peroxo complexes formed at the CNP surface and *in fine* the catalytic reactivity of CNPs by standard spectroscopy techniques.

## Supporting Information

Images of CNP8-$H_2O_2$ dispersions at different times after $H_2O_2$ addition (S1); Raman scattering of CN5 dispersions with and without $H_2O_2$ (S2); Materials and Methods (S3); X-ray photoelectron spectroscopy (XPS) (S4); Position of the absorbance excess peak as a function of $H_2O_2$ concentration (S5); Superposition of the normalized $\tilde{A}_{exc}(\lambda)$ for CNP5 and CNP8 under different physico-chemical conditions (S6); Absorbance excess for CNP23 and CNP28 (S7); Comparison between Day0 and Day30 absorbances for CNP8-$H_2O_2$ (S8).

## Acknowledgments

The authors would like to thank Ana Regina Simoes Sampaio for the UV-Visible spectrometry experiments on cerium oxide nanoparticle dispersions. ANR (Agence Nationale de la Recherche) and CGI (Commissariat à l'Investissement d'Avenir) are gratefully acknowledged for their financial support of this work through Labex SEAM (Science and Engineering for Advanced Materials and devices) ANR-10-LABX-0096 et ANR-18-IDEX-0001. We acknowledge the ImagoSeine facility (Jacques Monod Institute, Paris, France), and the France BioImaging infrastructure supported by the French National Research Agency (ANR-10-INBS-04, « Investments for the future »). This research was supported in part by the Agence Nationale de la Recherche under the contract ANR-15-CE18-0024-01 (ICONS), ANR-20-CE18-0022 (Stric-On) and by Solvay. We also ackowledge the travel support from Czech Ministry of Education, Youth and Sports (MSMT) Barrande Mobility program 8J23FR026.

## Graphical abstract

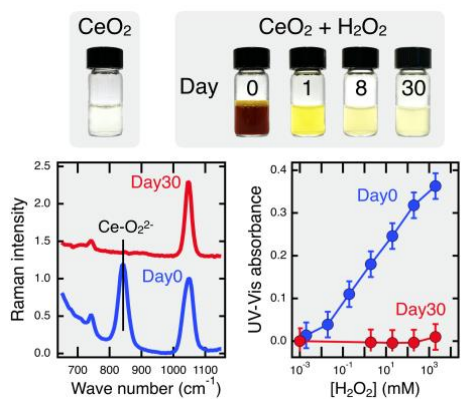

# Supporting Information

## Cerium oxide catalyzed disproportionation of hydrogen peroxide: a closer look at the reaction intermediate

**Giusy Finocchiaro[1,2]\*, Xiaohui Ju[3]\*, Braham Mezghrani[1] and Jean-François Berret[1]\***

[1]*Université Paris Cité, CNRS, Matière et systèmes complexes, 75013 Paris, France*
[2]*Institute of Photonics and Electronics of the Czech Academy of Sciences, Chaberská1014/57, 182 51 Prague, Czech Republic*
[3]*Center for Nanorobotics and Machine Intelligence, Department of Chemistry and Biochemistry, Mendel University in Brno, Zemedelska 1, 613 00 Brno, Czech Republic*

**Outline**

**S1** – Images of CNP8-$H_2O_2$ dispersions at different times after $H_2O_2$ addition
**S2** – Raman scattering of CNP5 dispersions with and without $H_2O_2$
**S3** – Materials and Methods
**S4** – X-ray photoelectron spectroscopy (XPS)
**S5** – Position of the absorbance excess peak as a function of $H_2O_2$ concentration
**S6** – Superposition of the normalized $\tilde{A}_{exc}(\lambda)$ for CNP5 and CNP8 under different physico-chemical conditions
**S7** – Absorbance excess for CNP23 and CNP28
**S8** – Comparison between Day0 and Day30 absorbances for CNP8-$H_2O_2$

**Corresponding authors**: Finocchiaro@ufe.cz, ju@mendelu.cz, jean-francois.berret@u-paris.fr
Version submitted Saturday, September 23, 23

## Supporting Information S1
### Images of CNP8-H₂O₂ dispersions at different times after H₂O₂ addition

Visual records of vials containing CNP8-H₂O₂ dispersions were monitored for 30 days after H₂O₂ addition. These results show that CNP8 dispersions can self-regenerate by catalyzing the disproportionation of H₂O₂.

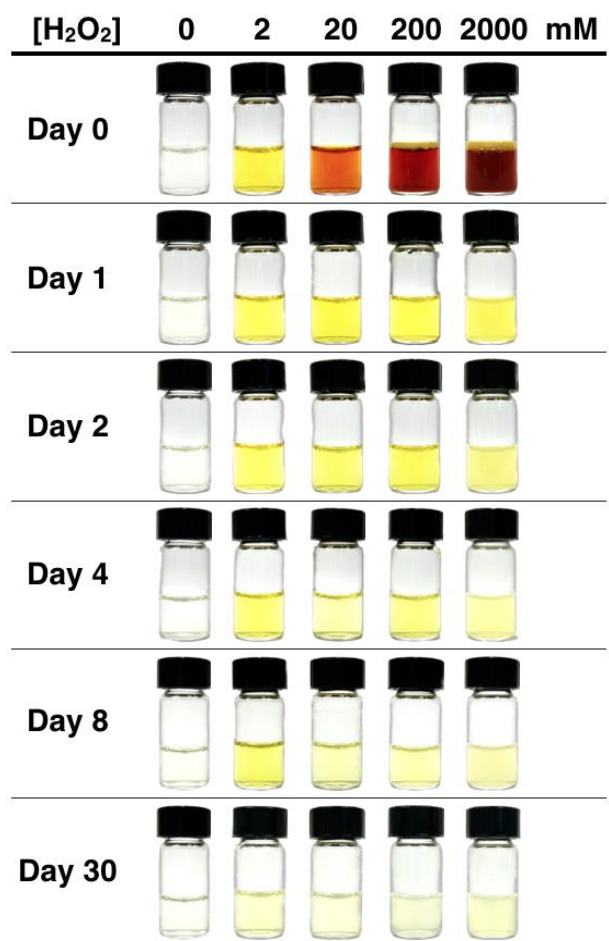

**Figure S1 :** Photographs of vials containing 8 g L⁻¹ CNP8 dispersions were captured following the addition of increasing concentrations of H₂O₂ (0, 2, 20, 200, and 2000 mM). The initial row displays the vials one hour post H₂O₂ introduction, while subsequent rows depict the same vials undergoing incubation over a 30-day period (T = 25 °C). These images illustrate the color changes of the CNP8 dispersions upon H₂O₂ addition, coupled with the completion of a redox cycle involving Ce(III) and Ce(IV) oxidation states.

## Supporting Information S2
**Raman scattering of CNP5 dispersions with and without H₂O₂**

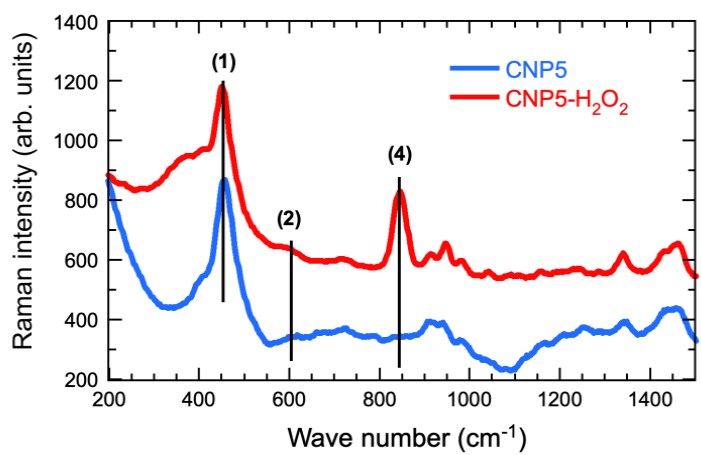

**Figure S2:** Raman spectra of CNP5 after $H_2O_2$ addition, recorded between 200 cm$^{-1}$ and 1500 cm$^{-1}$. The samples are in the liquid state and prepared under the same conditions as the CNP8 samples of Fig. 1d. The spectrum obtained with $H_2O_2$ highlights the presence of the band at 840 cm$^{-1}$ attributed to cerium-peroxo complexes Ce-O$_2$$^{2-}$.The Raman band at 460 cm$^{-1}$ **(1)** corresponds to the $F_{2g}$ mode of CeO$_2$ fluorite phase, whereas that at 590 cm$^{-1}$ **(2)** arises from defects in CeO$_2$ crystal structure, such as oxygen vacancies. Note the absence in both spectra of bands associated with nitrate ions, as found for CNP8.

## Supporting Information S3
**Materials and Methods**

### UV-Visible spectrophotometry
The absorbance of CNP aqueous dispersions (where CNP refers to CNP5, CNP8, CNP23 and CNP28) was measured with a UV-visible spectrophotometer (JASCO, V-630) equipped with a temperature controller. The CNP absorption spectra $A_{CNP}(\lambda)$ were recorded in the range $\lambda$ = 190 − 800 nm at room temperature (T = 25 °C). According to the Beer-Lambert law, $A_{CNP}(\lambda)$ is proportional to CNP concentration $c_{CNP}$: $A_{CNP}(\lambda) = \varepsilon_{CNP}(\lambda)lc_{CNP}$ where $\varepsilon_{CNP}(\lambda)$ are the absorptivity coefficient and $l$ the cell thickness. Given $l$ = 1 cm for a standard quartz SUPRASIL cell (Hellma, QS.10) and $\varepsilon_{CNP}$ = 25.2 L g$^{-1}$ cm$^{-1}$ at the characteristic 288 nm CeO$_2$ peak,[1-3] the concentration of the synthesized CNPs was evaluated using the Beer-Lambert law with accuracy better than 0.1% in the range 10$^{-3}$ - 10 g L$^{-1}$.

For the UV-visible absorption analysis, the concentration of CNP in the dispersions was reduced to 0.03 g L$^{-1}$ to record the characteristic peak of CeO$_2$ at $\lambda$ = 288 nm. To the same concentration of CNPs, different concentrations of $H_2O_2$ were added, ranging from 0.002 mM to 2000 mM. All the UV-visible absorption spectra ($\tilde{A}_{CNP}(\lambda)$ or $\tilde{A}_{CNP+H_2O_2}(\lambda)$) were treated in the same way: (i) subtracting the UV-vis absorption spectrum of the solvent ($A_{H_2O}(\lambda)$ or $A_{H_2O_2+H_2O}(\lambda)$) and (ii)





normalizing by the absorbance of CNP at λ = 288 nm ($A_{CNP}(\lambda = 288\ nm)$), according to the following equations:

(1) $\quad \tilde{A}_{CNP}(\lambda) = \frac{A_{CNP}(\lambda) - A_{H_2O}(\lambda)}{A_{CNP}(\lambda = 288\ nm)}$

(2) $\quad \tilde{A}_{CNP+H_2O_2}(\lambda) = \frac{A_{CNP+H_2O_2}(\lambda) - A_{H_2O_2+H_2O}(\lambda)}{A_{CNP}(\lambda = 288\ nm)}$

For the investigation of CNP8-$H_2O_2$ reaction over time, the samples were stored at 25 °C in hermetically closed vials to prevent the evaporation.

Finally, the absorbance excess is defined as the subtraction of the UV-vis absorption spectrum of CNP to the UV-vis absorption spectrum of CNP-$H_2O_2$ mixture, according to equation (3):

(3) $\quad \tilde{A}_{exc}(\lambda) = \tilde{A}_{CNP+H_2O_2}(\lambda) - \tilde{A}_{CNP}(\lambda)$

The maximum of $\tilde{A}_{CNP+H_2O_2}(\lambda)$ of CNPs at different sizes was determined for each $H_2O_2$ concentration and plotted as a function of the concentration of $H_2O_2$, to get a curve $\tilde{A}_{exc}([H_2O_2])$ that can be fitted with the Hill's equation (4), as described in the main text.

(4) $\quad \tilde{A}_{exc}([H_2O_2]) = \tilde{A}_{max} \frac{[H_2O_2]^\alpha}{c_0^\alpha + [H_2O_2]^\alpha}$

By dividing the CNP concentration (0.035 g L$^{-1}$) by the molecular weight of $CeO_2$ (172 g mol$^{-1}$), it is possible to calculate the molar concentration of cerium atoms [Ce] in the CNP dispersion, that is equal to around 0.20 mM. The ratio between the molar concentration of $H_2O_2$ required to reach the saturation (that can be calculated from the Hill's fitting of the $\tilde{A}_{exc}([H_2O_2])$ curve) and the molar concentration of cerium atoms provides the number of moles of $H_2O_2$ per mole of cerium. The same calculations were performed for all the investigated sizes of CNPs and gave an estimation of the number of moles of $H_2O_2$ that are required to reach the saturation, as reported in the table below.

| CNP size (nm) | [Ce] (mM) | [H$_2$O$_2$]$_{Sat}$ = 2$c_0$ (mM) | [H$_2$O$_2$]$_{Sat}$/[Ce] |
|---|---|---|---|
| 4.5 | 0.17 | 5.2 | 26 |
| 7.8 | 0.17 | 7.2 | 36 |
| 23 | 0.17 | 27.6 | 138 |
| 28 | 0.17 | 30.8 | 154 |

**Table S3:** Estimation of the number of moles of $H_2O_2$ per mole of cerium in the CNP dispersion.

**Transmission Electron Microscopy (TEM)**

Micrographs were taken with a Tecnai 12 TEM operating at 120 kV equipped with a 4-k camera OneView and the GMS3 software (Gatan). A drop of the CNP dispersion was deposited on





ultrathin carbon type-A 400 mesh copper grids (Ted Pella, Inc.). Micrographs were analyzed using ImageJ software for 250 particles. The particle size distribution was adjusted using a log-normal function of the form:

$$p(d, D, s) = 1/\sqrt{2\pi}\beta(s)d\, exp(-ln^2(d/D)/2\beta(s)^2),$$

where $D$ is the median diameter, and $\beta(s)$ is related to the size dispersity $s$ through the relation:

$$\beta(s) = \sqrt{\ln(1 + s^2)}.$$

$s$ is the ratio between the standard deviation and the average diameter. For $\beta < 0.4$, one has $\beta \cong s$ [4,5].

### X-ray Photoelectron Spectroscopy (XPS)

XPS analysis was performed using an Omicron Argus X-ray photoelectron spectrometer, equipped with a monochromated AlKα radiation source ($h\nu$ = 1486.6 eV) and a 280 W electron beam power. Experiments were performed on powdered samples, obtained either by heating (60° C for one day) and evaporation of solvent from the dispersion or by freeze-drying. In this way, we were able to check if heating modifies the surface defects associated with the oxygen vacancies and surface Ce(III). The emission of photoelectrons from the sample was analyzed at a takeoff angle of 45° under ultra-high vacuum conditions ($\leq 10^{-9}$ mbar). Spectra were carried out with a 100 eV pass energy for the survey scan and 20 eV pass energy for the C 1s, O 1s and Ce 3d regions. Binding energies were calibrated against the C 1s (C-C) binding energy at 284.8 eV and element peak intensities were corrected by Scofield factors. The peak areas were determined after subtraction of a Shirley background. The spectra were fitted using KolXPD software (kolibrik.net, s.r.o, Czech Republic) and applying a Gaussian/Lorentzian ratio G/L equal to 70/30. The measured core-level spectra of Ce 3d were fitted with five doublets corresponding to the Ce(III) and Ce(IV) states to evaluate the Ce oxidation state according to ref.[6].

### Raman scattering

For Raman scattering, CNP5 and CNP8 samples were prepared by mixing hydrogen peroxide ($H_2O_2$) solution at different concentrations ($0.03 - 30\%$) with CNP dispersion at 10 g $L^{-1}$ in a ratio $H_2O_2$:CNP8 of 1:4. Then, the samples were split into 2: one remains in the liquid state and the other was freeze-dried at -55 °C / 0.3 mBar for 24 hours. Both were stored at 4 °C and later studied by Raman scattering. The CNP8 dispersion without $H_2O_2$ was used as a reference. Specimens were analysed by Raman spectroscopy using a confocal microscope system LabRAM HR800 (HORIBA Scientific, Jobin-Yvon). The samples were illuminated through a 100x objective lens with 633 nm excitation from helium-neon (He-Ne) laser source at an incident power of 25 mW and with a spot diameter of approximately 1 μm. Raman spectra from these samples were collected over the range of $200 - 2400$ cm$^{-1}$. Each spectrum resulted from 3 scans, corresponding to a total collection time per spectrum of 30 s. The spectral resolution of all spectra was approximately 5 cm$^{-1}$.

## Supporting Information S4
### X-ray photoelectron spectroscopy (XPS)

To confirm the results obtained by Baldim et al. in 2018 [4], X-ray photoelectron spectroscopy experiments were performed on CNP5 (Fig. S4a) and CNP8 (Fig. S4b) powder samples. The results for the Ce(III) fraction are in agreement with the initial measurements. Data obtained for the various CNP samples by transmission electron microscopy, wide-angle X-Ray scattering and from X-ray photoelectron spectroscopy are displayed in Fig. S4c.

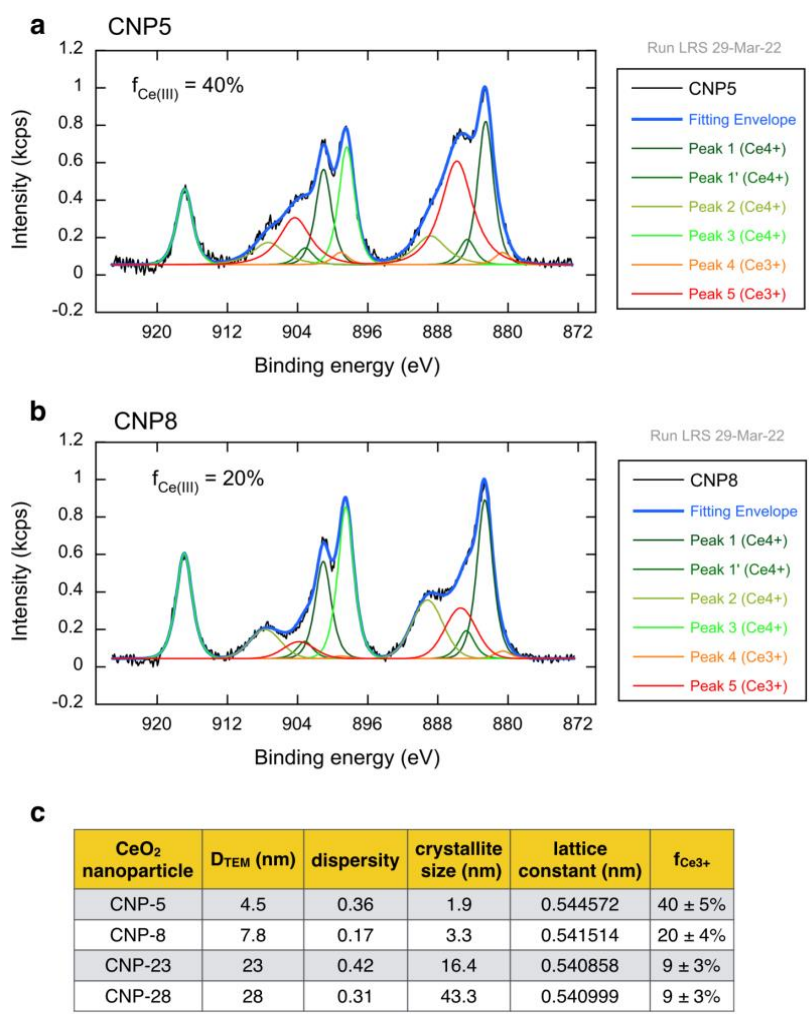

| CeO₂ nanoparticle | $D_{TEM}$ (nm) | dispersity | crystallite size (nm) | lattice constant (nm) | $f_{Ce3+}$ |
|---|---|---|---|---|---|
| CNP-5 | 4.5 | 0.36 | 1.9 | 0.544572 | 40 ± 5% |
| CNP-8 | 7.8 | 0.17 | 3.3 | 0.541514 | 20 ± 4% |
| CNP-23 | 23 | 0.42 | 16.4 | 0.540858 | 9 ± 3% |
| CNP-28 | 28 | 0.31 | 43.3 | 0.540999 | 9 ± 3% |

**Figure S4:** a) X-ray photoelectron spectroscopy of Ce3d core-level spectra for CNP5 powders. The continuous thick lines display the sum of the different peak contributions adjusted according to the model proposed by Y. Lykhach *et al.*[6]. The Ce(III) fraction for CNP5 sample was calculated from the integrated areas of the assigned peaks. b) Identical to a) for CNP8. c) Cerium oxide nanoparticle characteristics obtained from transmission electron microscopy, wide-angle X-Ray scattering [4] and from X-ray photoelectron spectroscopy [4,5,7].

## Supporting Information S5
**Position of the absorbance excess peak as a function of H2O2 concentration**

The position of the excess absorbance peak ($\lambda_{Max}$) associated with the Ce-O$_2^{2-}$ complexes is shown as a function of the H$_2$O$_2$ concentration in Fig. S5. The CNP-H$_2$O$_2$ dispersions exhibit a $\lambda_{Max}$ ranging from 360 nm to 380 nm. To simplify, in the main text we refer to the peak position at the value $\lambda_{Max}$ = 365 nm. There are two outliers, which are found from CNP23 and CNP28 at [H$_2$O$_2$] = 200 mM and 2000 mM. These latter data are not considered in the analysis proposed in the main text (see also S7 for details).

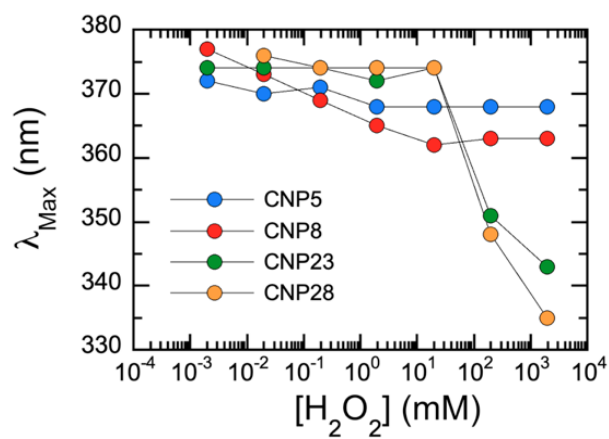

**Figure S5:** Position of the excess absorbance peak ($\lambda_{Max}$) versus H$_2$O$_2$ concentration for CNP5, CNP8, CNP23 and CNP28.

## Supporting Information S6
**Superposition of the normalized $\widetilde{A}_{exc}(\lambda)$ for CNP5 and CNP8 under different physico-chemical conditions**

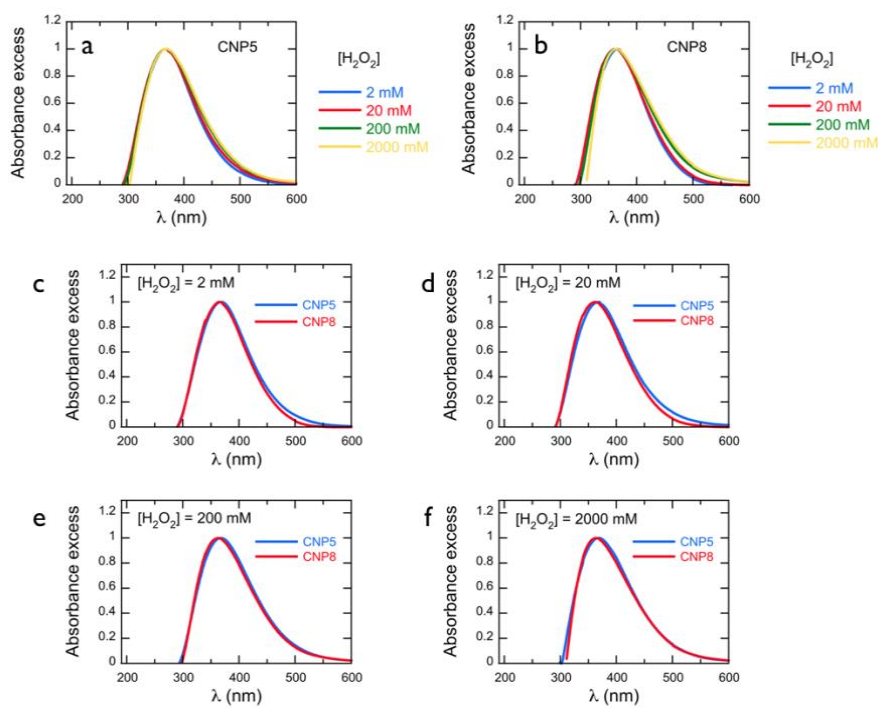

**Figure S6:** a) Effect of H₂O₂ concentration on the position and shape of the 365 nm absorbance peak for CNP5-H₂O₂ dispersions (T = 25 °C). Data have been normalized to the maximum absorbance value at $\lambda_{Max}$. b) Identical to a) for CNP8-H₂O₂. c) Comparison of absorbance peaks for CNP5-H₂O₂ and CNP8-H₂O₂ dispersions for [H₂O₂] = 2 mM. Data have been normalized to the maximum absorbance value at $\lambda_{Max}$. d,e,f) Identical to c) for [H₂O₂] = 20, 200 and 2000 mM, respectively. The results show a good overlap of the absorbance excess bands, indicating a common underlying phenomenon.

## Supporting Information S7
**Absorbance excess for CNP23 and CNP28**

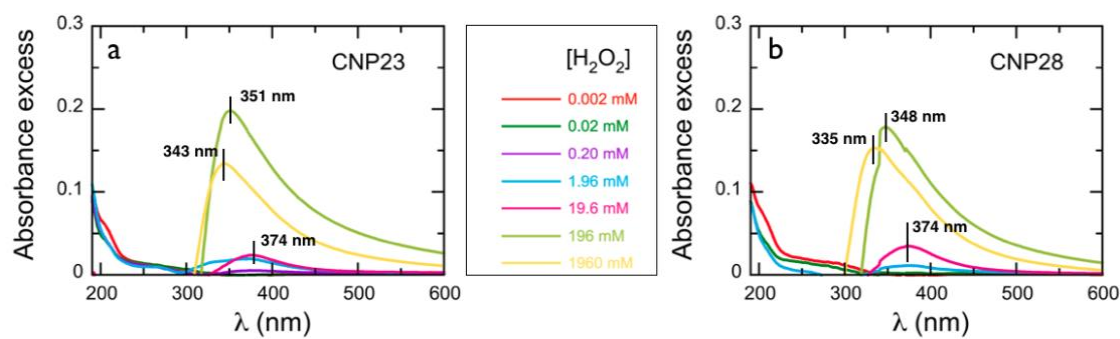

**Figure S7:** a) Absorbance excess $\tilde{A}_{exc}(\lambda)$ versus wavelength for CNP23-$H_2O_2$ (T = 25 °C). b) Same as in a) for CNP28-$H_2O_2$. Only spectra obtained for $H_2O_2$ concentrations below 20 mM show an absorbance band centered around 370 nm, in agreement with the data on CNP5 and CNP8.

## Supporting Information S8
**Comparison between Day0 and Day30 absorbances for CNP8-$H_2O_2$**

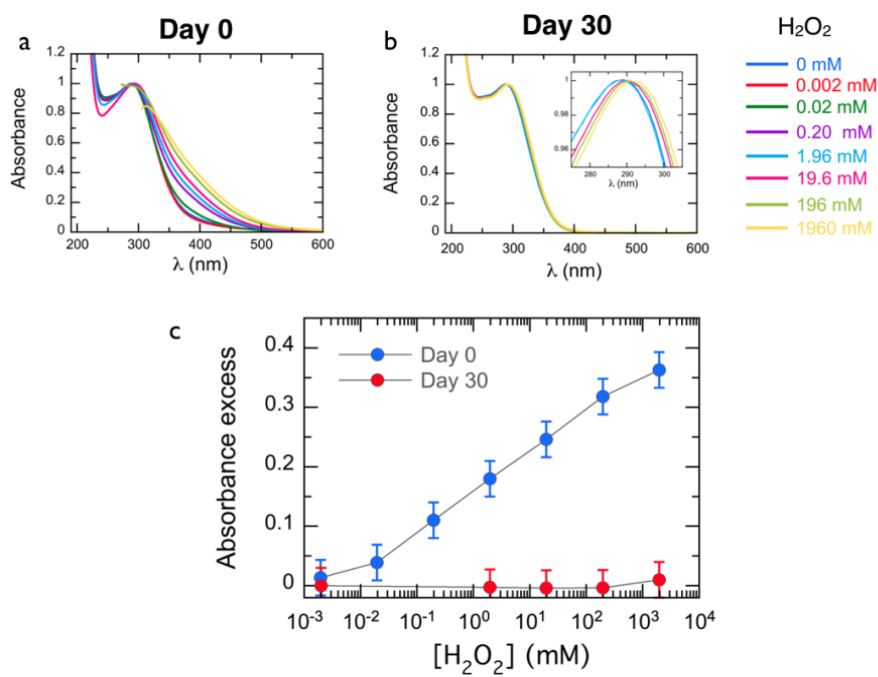

**Figure S8:** a) UV-Visible absorbance spectra of 0.03 g $L^{-1}$ of CNP8 with increasing amount of $H_2O_2$ from 0.002 to 2000 mM. The figure here is identical to Fig. 2f. The absorbance measurements were taken one hour after the addition of hydrogen peroxide. b) Identical to a) for CNP8-$H_2O_2$ dispersions which have been kept at room temperature for one month. Inset: Zoom in on the area showing maximum absorbance for the different $H_2O_2$ concentrations. c) Absorbance excess $\tilde{A}_{exc}(\lambda) = \tilde{A}_{CNP+H_2O_2} - \tilde{A}_{CNP}$ at the wavelength $\lambda$ = 365 nm one hour and one month after addition of increasing amounts of $H_2O_2$ to the CNP8 dispersion.